\pdfoutput=1
\documentclass[11pt]{article}
\usepackage{authblk}

\usepackage{ACL2023}

\usepackage{times}
\usepackage{latexsym}
\usepackage{graphicx}
\usepackage{multirow}
\usepackage{booktabs}
\usepackage{amsmath}
\usepackage{bm}

\usepackage[T1]{fontenc}

\usepackage[utf8]{inputenc}

\usepackage{microtype}

\usepackage{inconsolata}
\title{Towards Alleviating the Object Bias in Prompt Tuning-based  Factual Knowledge Extraction}
\author[$\dagger$]{\textbf{Yuhang Wang}}
\author[$\ddagger$]{\textbf{Dongyuan Lu}}
\author[$\dagger$]{\textbf{Chao Kong}}
\author[$\dagger$\thanks{Corresponding author}]{\textbf{Jitao Sang}}
\affil[$\dagger$]{Beijing Key Lab of Traffic Data Analysis and Mining \authorcr
          Beijing Jiaotong University, Beijing, China \authorcr
          \{yhangwang, kongchao, jtsang\}@bjtu.edu.cn}
\affil[$\ddagger$]{School of Information Technology and Management \authorcr
          University of International Business and Economics, Beijing, China\authorcr
          ludy@uibe.edu.cn}

\begin{document}
\maketitle
\begin{abstract}
Many works employed prompt tuning methods
to automatically optimize prompt queries and 
extract the factual knowledge 
stored in Pretrained Language Models. 
In this paper, 
we observe that the optimized prompts, 
including discrete prompts and continuous prompts,
exhibit undesirable object bias.
To handle this problem, 
we propose a novel prompt tuning method called MeCoD\footnote{https://github.com/astrodrew/MeCoD}.
consisting of three modules: 
Prompt Encoder, Object Equalization and
Biased Object Obstruction.
Experimental results show that 
MeCoD can significantly 
reduce the object bias and 
at the same time 
improve accuracy of factual knowledge extraction.
\end{abstract}

\section{Introduction}
Pretrained language models (PLMs) have become a standard practice in NLP and achieved strong performance
on many downstream tasks \citep{Qiu2020PretrainedMF,Liu2021PretrainPA}.
A recognized reason why PLMs are so powerful is the knowledge
learned from a large amount of public corpus \citep{Liu2019LinguisticKA}.
Recently, researchers have taken interest in 
measuring and extracting the factual knowledge in PLMs. 
\citet{Petroni2019LanguageMA} first formally proposed the LAMA benchmark,
which employs hand-crafted prompts to retrieve factual knowledge in the form of
< subject, relation, object > triples.
For example, 
regarding a factual knowledge triple 
< \emph{Douglas Adams}, \emph{native language}, \emph{English} >,
LAMA can query PLMs with ``\emph{The native language of Douglas Adams is [MASK]}'' 
to extract the native language of \emph{Douglas Adams},
where ``\emph{The native language of [X] is [MASK]}'' 
is a manual prompt for the relation ``\emph{native language}''
and ``\emph{[MASK]}'' is a placeholder for the object to predict.
\par In order to extract factual knowledge more effectively, 
many works take a step toward automatically tuning prompts 
with additional training set.
\citet{Shin2020ElicitingKF} proposed AutoPrompt to generate discrete prompts automatically 
based on gradient optimization by maximizing the expected likelihood of the ground truth object.
Instead of searching discrete prompts, 
a more flexible research line is tuning continuous prompts directly in the input embedding space. 
For example,
\citet{Liu2021GPTUT} proposed P-tuning to optimize a continuous prompt for each factual relation, 
and achieved SOTA performance.
\citet{Li2021PrefixTuningOC} proposed a semi-automatic method called Prefix-tuning 
to learn a prefix to add to manual prompts. 
\citet{Newman2022PAdaptersRE} applied Prefix-tuning 
to improve the robustness of factual knowledge extraction.
\begin{figure}[tbp] 
  \includegraphics[width=\linewidth]{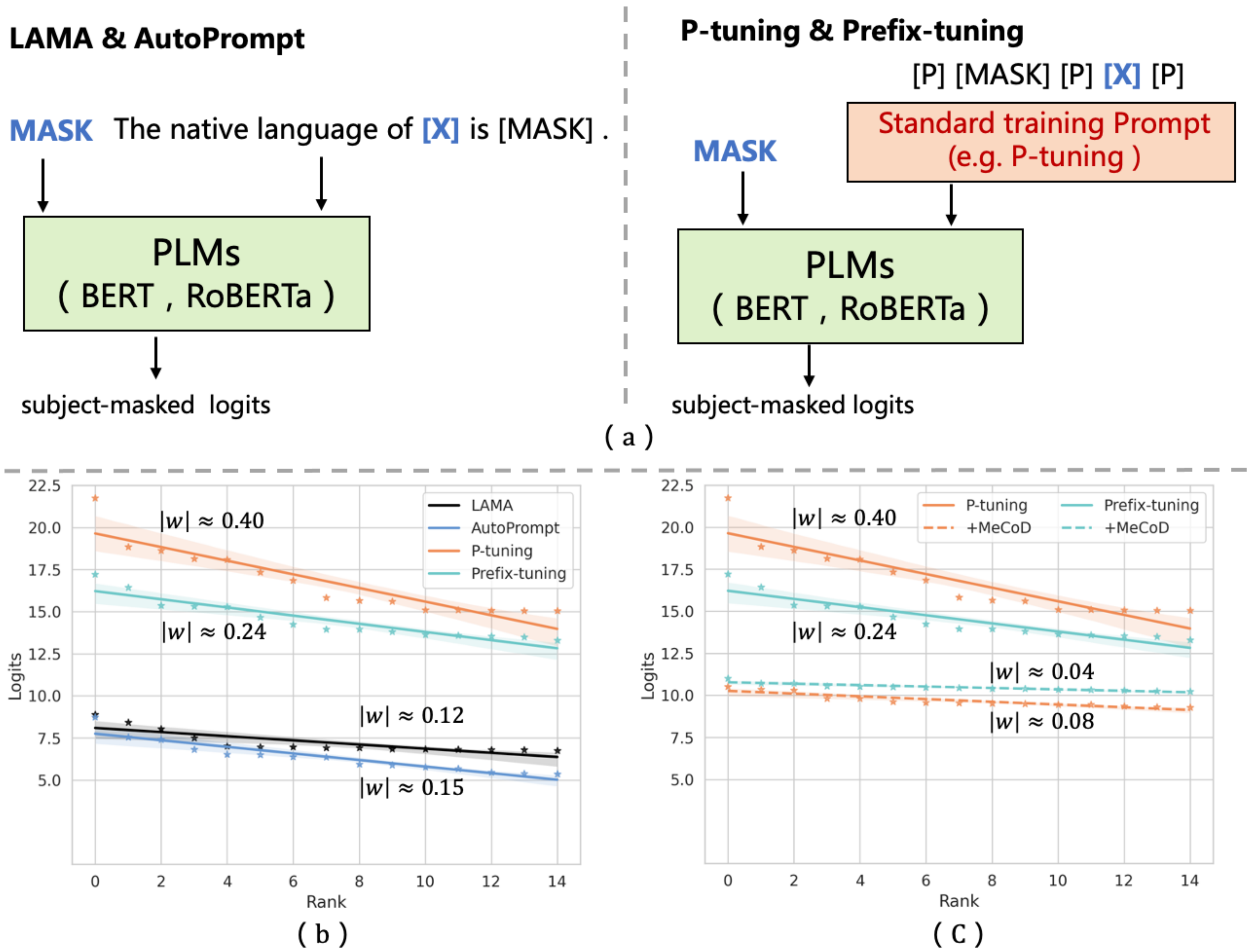}
  \centering
  \caption{
           \emph{Object bias} in different prompt-based knowledge extraction methods: 
           LAMA, AutoPrompt, Prefix-tuning and P-tuning. 
           (a) demonstrates how to construct subject-masked prompts. 
           (b), (c) show the derived logits of top-retrieved objects for original and
            our proposed prompt-tuning methods, respectively
           } 
  \label{fig:demo}
  \vspace{-0.3cm}
\end{figure}
\par  Although the above prompt tuning methods achieve good performance,
we discuss in this paper that they suffer from 
severe \emph{object bias} problem.
\begin{table*}[t]
    \centering
    \begin{tabular}{lll}
      \toprule
      Extraction Method &  \multicolumn{2}{c}{Prompt Template}  \\ 
      \midrule
      \multirow{2}{*}{LAMA} & \emph{Original} & The native language of Pierre Messmer is \textbf{[MASK]} .\\
                            & \emph{Subject-masked} & The native language of [MASK] is \textbf{[MASK]} . \\
      \midrule
      \multirow{2}{*}{AutoPrompt} & \emph{Original} & Pierre Messmer [T] \textbf{[MASK]} . \\
                                  & \emph{Subject-masked}  & [MASK] [T] \textbf{[MASK]} \\
      \midrule
      \multirow{2}{*}{Prefix-tuning} & \emph{Original} & [P] The native language of Pierre Messmer is \textbf{[MASK]} . \\
                                     & \emph{Subject-masked} & [P] The native language of MASK is \textbf{[MASK]} . \\
      \midrule
      \multirow{2}{*}{P-tuning} & \emph{Original} & [P] \textbf{[MASK]} [P] Pierre Messmer [P] \\
                                & \emph{Subject-masked} & [P] \textbf{[MASK]} [P] [MASK] [P] \\
      \bottomrule
    \end{tabular}
    \caption{Example of original and subject-masked prompt templates for relation P103. 
            ``[T]'',``[P]'' indicate discrete and continuous optimizable prompt token,
            respectively.
             The number of [P] and [T] can be customized. 
             ``[MASK]'' in bold is the placeholder for the object to predict.}
    \label{tab:subject-mask-input}
\end{table*}
As illustrated in Figure \ref{fig:demo}(a), 
we construct subject-masked prompts for different prompt-based knowledge extraction methods.  
Example prompts for relation P103 are illustrated in Table \ref{tab:subject-mask-input}.
We conduct experiments on LAMA benchmark \citep{Petroni2019LanguageMA} which consists of 41 fact relations. 
Figure \ref{fig:demo}(b) shows the derived logits of top-k retrieved objects in descending order. 
Since the subject is masked in the issued prompt template, 
no context is provided and an even logit distribution for different object candidates is expected.
Taking the fact < \emph{Douglas Adams}, \emph{native language}, \emph{English} > for example,
objects like ``French'', ``English'' and
``Russian'' should be treated equally
when \emph{Douglas Adams} is masked.
However,
we observe non-trivial slopes ($|w|$ in Figure \ref{fig:demo} (b)) of the regression lines 
in the 4 examined knowledge extraction methods,
i.e., they all exhibit bias towards specific objects. 
Notably, the 3 prompt-tuning methods of AutoPrompt, P-tuning and Prefix-tuning, 
have more inclined slopes and thus exhibit more severe object bias. 
More object bias measurement results are available at Section \ref{sec:Object Bias Measurement}.
Given the observed object bias in prompt tuning methods, 
we further design analysis experiments 
and find the negative influence of object bias 
on knowledge extraction accuracy (detailed in Section \ref{sec:Influence on Knowledge Extraction}). 
This motivates us to develop solutions to both alleviate the object bias 
problem and contribute to more accurate factual knowledge extraction.
\par In this paper,
to address the object bias problem in prompt-tuning stage, 
we propose \textbf{MeCoD} 
(\textbf{M}aximum \textbf{e}ntropy and \textbf{C}ontrastive learning 
for \textbf{o}bject \textbf{D}ebiasing) 
towards unbiased factual knowledge extraction~\footnote{~\small{
Since continuous prompts are more effective and widely adopted, 
MeCoD is designed to improve continuous prompt tuning methods, 
e.g., P-tuning, Prefix-tuning.}}. 
The basic idea is deriving equalized object predictions with subject-masked prompt, 
and at the same time disencouraging the biased objects with original prompt. 
These goals are realized by a maximum entropy-based Object Equalization module 
and contrastive learning-based Biased Object Obstruction module, respectively.
Figure \ref{fig:demo} (c) illustrates 
the intuitive effect of object bias alleviation.
\\\noindent\textbf{Contributions.} 
We summarize the main contributions of this paper as follows:
\begin{itemize} 
  \item We position the object bias problem in 
  prompt tuning-based factual knowledge extraction. 
  The influence of object bias on knowledge extraction accuracy is also discussed.
  \item We propose an object debiasing method at the prompt tuning stage
  to alleviate the object bias and improve accuracy of factual knowledge extraction. 
  \item The effectiveness of the proposed method is validated
  with sufficient qualitative and quantitative experiments.
\end{itemize}

\section{Data Analysis \label{sec:data analysis}} 
\textbf{Object Bias Definition.}
Factual knowledge can be represented in form of
<subject, relation, object> triples.
\emph{Object bias} in factual knowledge extraction
refers to the phenomenon that 
the pretrained language model with prompts
retrieves object candidates unequally when subject is not assigned,
e.g., prefering ``French'' to ``English'' 
in the prediction of person's native language 
when the person is not specified.
\subsection{Object Bias Measurement. \label{sec:Object Bias Measurement}}
Object bias inherently considers the uncertainty of retrieved objects with
subject-masked prompt queries. We thus employ entropy \citep{shannon1948mathematical} 
in this work to measure object bias.
Specifically, we define \emph{object bias entropy} 
in terms of the relation $R$ as:
\begin{equation}
   \begin{aligned}
    {\rm entropy}(R) &= -\sum_{i=1}^k p_R(i){\rm log}_2(p_R(i)) ,
   \end{aligned}
\label{equ:entropy}
\end{equation}
where $p_R$ is obtained by 
selecting top-k subject-masked logit values and normalizing with softmax function,
$k$ denotes the number of logit values used to calculate entropy.
In our subsequent analysis, we set $k$ to 10, 
and ${\rm entropy}(R)$ will achieve a maximum value of about 2.305 
when the object logits are equal. 
The smaller the value, 
the more significant the object bias.
\par We measure the 4 typical factual knowledge extraction methods
on the LAMA benchmark according to Eqn.\ref{equ:entropy}.
\begin{table}[t]
  \centering
  \tabcolsep=0.08cm
  \begin{tabular}{lcc}
    \toprule 
     Method &  Entropy & Comparison with 2.305\\ 
    \midrule
     LAMA &  2.077 & -9\% \\
     AutoPrompt &  1.901 & -17\%\\
     P-tuning &  1.754 & -23\% \\
     Prefix-tuning & 2.002 & -13\%\\
    \bottomrule
  \end{tabular} 
  \caption{The averaged object bias entropy over 41 relations of different knowledge 
  extraction methods on LAMA benchmark.}
  \label{tab:entropy}
\end{table}
The averaged result over 41 relations is shown in Table \ref{tab:entropy}.
It is easy to find that 
the observation in the form of object bias entropy 
is consistent with that of slope in Figure \ref{fig:demo}:
(1) The 4 methods all exhibit object bias,
the entropy values noticeable decrease from 2.305 by 9\%, 17\%, 23\% and 13\%, 
respectively. 
(2) The object bias entropy of prompt tuning methods, 
including AutoPrompt, Prefix-tuning and P-tuning,
is more smaller than that of manual prompts, LAMA.
This observation further demonstrates that 
prompt tuning methods suffer more serious object bias than manual prompts.
Note that the object bias entropy of
Prefix tuning falls in between
that of manual prompts, LAMA
and full-automatic prompts, AutoPrompt and P-tuning.
A possible reason is that
the manual template in the Preﬁx-tuning 
limits the learning of object bias.
\subsection{Influence on Knowledge Extraction \label{sec:Influence on Knowledge Extraction}}
In order to investigate the influence of object bias 
on knowledge extraction,
we compare the 
retrieved object candidates from
original prompts 
and subject-masked prompts.
Specifically,
we examine the rank of ground-truth object 
in the retrieved object lists and 
calculate Pearson correlation coefficient 
between the rank corresponding to 
the original prompt and the subject-masked prompt. 
\par According to the result on all testing samples,
we find that the correlations of prompt tuning methods
are higher than that of LAMA (Table \ref{tab: negative impact}).
This illustrates the prediction results of prompt tuning methods 
are influenced more significantly by object bias than that of manual prompts.
Furthermore, 
comparing between the results of all testing samples 
and incorrectly predicted samples, 
it is easy to observe that correlation coefficient of LAMA remains 
almost unchanged,
while those of the 3 prompt tuning methods 
increase obviously in incorrectly predicted samples.
This suggests that
some of the incorrect predictions
can be attributed to the bias towards specific objects, 
and motivates us to address the object bias problem to further
improve knowledge extraction accuracy.
\begin{table}[t]
    \centering
    \begin{tabular}{lcc}
      \toprule 
       Method & All   & Incorrectly predicted\\ 
      \midrule
      LAMA & 0.260 & 0.258 \\
      AutoPrompt & 0.380 & 0.416 \\
      P-tuning & 0.432 & 0.479 \\
      Prefix-tuning & 0.374 & 0.394 \\
      \bottomrule
    \end{tabular}
    \caption{The Pearson correlation coefficient 
     between the rank corresponding to
     original prompt and subject-masked prompt.}

     \label{tab: negative impact} 
  \end{table}
\begin{table}[t]
  \centering
  \tabcolsep=0.2cm
  \begin{tabular}{lcc}
    \toprule 
     Dateset & P@1 & Entropy\\ 
    \midrule
     Original & 49.413 & 1.754\\
     Undersampled & 48.776 & 1.924\\ 
    \bottomrule
  \end{tabular}
  \caption{
    Results for P-tuning fit to original dataset
    and undersampled dataset.}
  \label{tab: balanced dataset ptuning} 
\end{table}
\begin{figure*}[t] 
  \includegraphics[width=\linewidth]{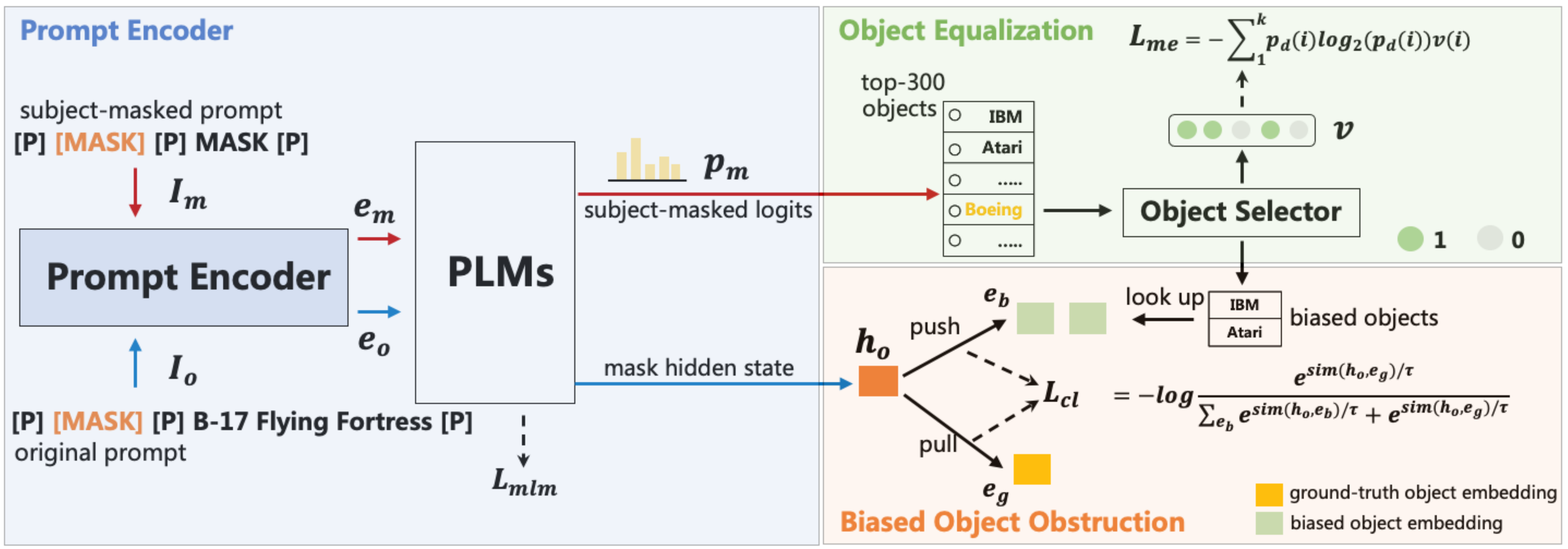}
  \centering
  \caption{Overall architechture of the proposed MeCoD.} 
  \label{fig:framework}
\end{figure*}

\subsection{Undersampling-based Preliminary Attempts on Object Debiasing.}
The above observations demonstrate the necessity for object debiasing.
Object bias is attributed to 
both the pretraining stage of PLMs and prompt tuning stage. 
The observed object bias of LAMA mainly origins from the pre-training stage.
While some pilot studies~\citep{Guo2022AutoDebiasDM} 
are devoted to reducing bias in the pre-trained language models, 
we found in the above analysis that the object bias 
at prompt-tuning stage is more severe than 
that at the pre-training stage,
and thus focus on object debiasing at the prompt tuning stage 
in this paper.
\par The straightfoward cause of object bias 
at the prompt tuning stage is 
imbalanced training data for optimizing prompts.
Take P-tuning as an example,
we make a preliminary attempt by
retraining it with undersampled balanced training set.
In order not to excessively reduce the number of samples,
we first group the training samples according to objects,
and then randomly undersample the two groups 
with the largest number of samples. 
In this case,
their numbers are consistent with 
the number of the third largest group.
Table \ref{tab: balanced dataset ptuning} 
summarizes the performance of P-tuning 
trained with different training sets.
\par We observe that object bias is alleviated on 
undersampled training set,
and this validates the attribution of imbalanced tuning set 
in deriving object bias.
However, we find that the P@1 drops clearly
due to insufficient use of data,
that is, accuracy is sacrificed for debiasing.
This inspires us to design an effective prompt tuning method,
instead of simply balancing training data,
to alleviate object bias as well as improve accuracy performence.
\section{Methodology \label{sec:method}}
We present the overall framework, MeCoD, 
as illustrated in Figure \ref{fig:framework}. 
The basic idea is to improve prompt encoder so that 
issuing the resultant embeddings to popular PLMs no longer exhibits object bias.
The goals are two-fold: 
(1) Objects should have equal opportunities to be extracted from PLMs
by subject-masked prompt; 
(2) The biased objects should be prevented from being extracted by 
the original prompt with specified subject.
Correspondingly, MeCoD includes three modules.
The first module is Prompt Encoder to be optimized,
which takes original and subject-masked prompts as inputs
and issues the resultant embeddings to PLMs 
to obtain mask hidden state $h_o$ and subject-masked logits $p_m$ respectively
(see Section \ref{sec:Prompt Encoder}).
The second module is Object Equalization, 
which takes subject-masked logits $p_m$ as input 
and forces model to treat objects equally,
when the subject is masked (see Section \ref{sec:Maximum Entropy}). 
The third module is Biased Object Obstruction 
which further prevents the biased objects
from being extracted
by forcing the mask hidden state $h_o$ away from
biased object embeddings
and close to ground-truth object embedding
(see Section \ref{sec:Contrastive Learning}).
We will take P-tuning as an example to elaborate the details of each module below.
\subsection{Prompt Encoder \label{sec:Prompt Encoder}}
In this module, 
we first construct subject-masked input by 
replacing ``\emph{[X]}'' with ``\emph{[MASK]}'' ,
as shown in Table \ref{tab:subject-mask-input}.
The number of ``\emph{[MASK]}'' is set 
as the number of tokenized subject.
As shown in Figure \ref{fig:framework} (left),
given original prompt $I_o$ and subject-masked prompt $I_m$,
we use Prompt Encoder 
to get input embeddings $e_{o}$ and $e_{m}$ for PLMs.
Then, 
mask hidden state $h_{o}$,
subject-masked logits $p_{m}$
and the MLM (Masked Language Modeling) loss $\mathcal{L}_{mlm}$,
can be obtained from PLMs as follows: 
\begin{equation}
  \begin{aligned}
    h_{o}  &= {\rm PLMs}(e_{o}), 
    h_{m}   = {\rm PLMs}(e_{m}),\\
    p_{o}  &= {\rm MLM\text{-}head}(h_{o}),\\
    p_{m}  &= {\rm MLM\text{-}head}(h_{m}),\\
    \mathcal{L}_{mlm} &= -\frac{1}{N} \sum_{i=1}^{N}y_{i} \log_2(p_{o_{i}}),
  \end{aligned}
\label{equ:mlm output}
\end{equation}
where $y_{i}$ denotes the ground truth.
$p_{m}$ will be used to equalize the objects
with respect to subject-masked prompt in Section \ref{sec:Maximum Entropy}.
Both $h_{o}$ and $p_{m}$ will be employed to obstruct the influence
of biased objects
in Section \ref{sec:Contrastive Learning}.
\subsection{Object Equalization \label{sec:Maximum Entropy}}
According to data analysis in Section \ref{sec:data analysis},
we consider that 
the probabilities of object candidates 
should be equalized   
when issuing the subject-masked prompt to PLMs.
In this subsection, 
we will introduce a method based on maximum entropy to force
objects to be treated equally
when subject is not given.
Note that only the fact-related objects 
need to be considered, 
e.g., regarding relation P103 (native language), 
objects with risk to bias the prediction results are like 
``English'', ``French'' instead of ``apple''.
To filter out the unrelated objects, 
we first sort subject-masked logits $p_{m}$
with descending order to get $p_d$, 
and empirically reserve the top-300 objects $c_o$.
Then, we further employ a linear layer
as a binary classifier named Object Selector 
to identify the objects to be equalized.
Specifically,
the object selector takes the object embeddings as input
and returns a binary vector 
$v \in \{0,1\}^{300} $ 
with gumbel softmax \citep{Jang2017CategoricalRW} as follows:
\begin{equation}
  \begin{aligned}
   v = {\rm gumbel\text{-}softmax}({\rm Linear}(E(c_o))),
  \end{aligned}
\label{equ:selector}
\end{equation}
where $E$ denotes embedding layer of PLMs.
The object sets corresponding to $v(i)=1, i=1,2,...,300$ 
are selected to be equalized.
Finally,
we construct the loss $\mathcal{L}_{me}$ based on Maximum Entropy:
\begin{equation}
  \begin{aligned}
    \mathcal{L}_{me} &= -\sum_{i=1}^{300} p_{d}(i){\rm log_2}(p_{d}(i))v(i) .
  \end{aligned}
\label{equ:loss_entropy}
\end{equation}

\subsection{Biased Object Obstruction\label{sec:Contrastive Learning}}
\begin{table*}[t]
  \centering
  \tabcolsep=0.17cm
  \begin{tabular}{lllllll}
    \toprule 
    \multirow{2}{*}{Method} & \multicolumn{3}{c}{BERT-base} & \multicolumn{3}{c}{RoBERTa-base} \\ 
    \cmidrule{2-7}
    & P@1 & MRR & Entropy &  P@1 & MRR & Entropy \\ 
    \midrule
     LAMA	& 29.641 & 39.312	& 2.077 & 15.206 & 22.791	& 1.972 \\
    \midrule
     Prefix-tuning & 47.472 & 57.522 & 2.002 & 45.828 & 56.473 & 1.904\\
     + Undersampling & 42.254 & 52.914 & 2.159 & 40.742 & 52.086 & 2.100\\ 
     + MeCoD  & 48.329(2\% $\uparrow$) & 58.553 & \textbf{2.145}(7\% $\uparrow$) &
     46.281(1\% $\uparrow$) & 57.072 & \textbf{2.298}(16\% $\uparrow$) \\
     \midrule
     P-tuning & 49.413 & 59.419 & 1.754 & 44.828 & 54.955 & 1.655 \\
     + Undersampling  & 48.776 & 58.696 & 1.926 & 42.498 & 53.203 & 2.020 \\
     + MeCoD  & \textbf{50.438}(2\% $\uparrow$) & \textbf{60.335} & 2.141(22\% $\uparrow$) & 
     \textbf{46.813}(4\% $\uparrow$) & \textbf{57.474} & 2.109(27\% $\uparrow$) \\
    \bottomrule
  \end{tabular}
  \caption{Results on the LAMA benchmark using the BERT-base-cased and RoBERTa-base model.}
  \label{tab: result}  
\end{table*}

This module further 
reduce the probability of retrieving biased objects 
when issuing prompt with specified subjects,
and we introduce a module based on contrastive learning.  
The key idea is to simultaneously minimize the representation gap 
between ``\emph{[MASK]}'' and ground-truth object
and maximize that
between ``\emph{[MASK]}'' and irrelevant biased objects.
Specifically,
we regard the objects corresponding to $v(i)=1, i=1,2,...,300$ 
as biased objects except for the ground-truth object.
We formalize it as a contrastive learning problem 
and propose to minimize the following loss \citep{Oord2018RepresentationLW}:
\begin{equation}
  \small
  \mathcal{L}_{cl} = -{\rm log} \frac{e^{{\rm sim}(h_{o},e_{g})/\tau}}{\sum_{\substack{e_{b}}}e^{{\rm sim}(h_{o},e_{b})/\tau} + e^{{\rm sim}(h_{o},e_{g})/\tau} }, 
  \label{equ:loss_cl}
\end{equation} 
where sim(·) calculates the consine similarity of different representations,
$e_g$ and $e_{b}$ denote the word embeddings of 
ground-truth object and biased objects, respectively.
$\tau$ is the temperature, controlling the difficulty of distinguishing between positive and negative samples.
Intuitively, the contrastive loss forces the model
to push the mask hidden state away from the embeddings of biased objects,
and pull it to the ground-truth object embedding.
\par During training, the model 
is optimized by jointly minimizing loss $\mathcal{L}$ as follows:
\begin{equation}
  \begin{aligned}
   \mathcal{L}       &= \mathcal{L}_{mlm} - \lambda_{1} \mathcal{L}_{me} + \lambda_{2} \mathcal{L}_{cl} , \\
  \end{aligned}
\label{equ:loss}
\end{equation}
where $\lambda_{1}$ and $\lambda_{2}$ are the coefficients to balance the three training losses.

\begin{table*}[t]
  \centering
  \begin{tabular}{lcccc}
    \toprule 
    \multirow{2}{*}{Method} & \multicolumn{2}{c}{BERT-base} & \multicolumn{2}{c}{RoBERTa-base} \\ 
    \cmidrule{2-5} 
     & P@1  & Entropy &  P@1 & Entropy \\ 
    \midrule
     P-tuning + MeCoD & \textbf{50.438} & 2.141 & 
     \textbf{46.813}  & \textbf{2.109} \\
     w/o Object Equalization  & 50.301 &1.808 & 46.471 &  1.661  \\
     w/o Biased Object Obstruction  &50.317  &\textbf{2.212}  & 46.625 & 1.984  \\
    \bottomrule
  \end{tabular}
  \caption{Ablation study.}
  \label{tab: ablation study} 
\end{table*}
\section{Experiments \label{sec:exp}}
\subsection{Experimental Settings \label{sec:exp setup}}
\textbf{Datasets.}
We adopt the LAMA-TREx \citep{Petroni2019LanguageMA}
as the main testing set 
which consists of 41 Wikidata relations and altogether 29,500 testing triples.
Besides,
we also evaluate our method on WIKI-UNI \citep{cao2021knowledgeable}
which is constructed 
to ensure each object to appear the same times for each relation.
As for training,
we use the data collected by \citet{Shin2020ElicitingKF}
which contains 800 training samples and 200 developing samples for each fact relation.\\
\textbf{Evaluation metrics and baselines.}
In addition to object bias entropy,
we also evaluate the performance on knowledge extraction
with metrics of precision-at-1 (P@1) and mean reciprocal rank (MRR).
We report average performance over 41 relations. 
In order to evaluate the effectiveness of the proposed MeCoD,
we implement the LAMA, Prefix-tuning, P-tuning and Undersampling-based solutions (see Section \ref{sec:data analysis}) as baselines.
Specifically,
LAMA provides hand-crafted prompts that
are less object-biased than prompt tuning methods.
P-tuning and Prefix-tuning are the representatives of continuous prompt,
which are more effective and widely used.\\
\textbf{Implementation details.}
For PLMs in our experiments,
we investigate BERT \citep{Devlin2019BERTPO} and 
RoBERTa \citep{Liu2019RoBERTaAR}
\footnote{We use the the implementations of huggingface.
https://huggingface.co}.
The prompt encoder is initialized by standard training prompt encoder,
e.g.,P-tuning based on LSTM \citep{Shi2015ConvolutionalLN}.
We use the Adam optimizer \citep{kingma2014adam} with its default conﬁguration. 
For gradient training, we fix parameters of PLMs, 
and set the learning rate to 1e-5
to jointly optimize Prompt Encoder and Object Selector.
We set $\lambda_{1}$ and $\lambda_{2}$ to 0.2, 0.1 respectively.
As for training time,
take MeCoD on P-tuning as an example,
it spends about 40 minutes for each relation on
1 GPU device of RTX A4000.
\subsection{Quantitative Results}
Table \ref{tab: result} shows the performance of each method
on the two selected PLMs.
Briefly, 
MeCoD outperforms the baselines 
on both object bias and knowledge extraction performance.
Take P-tuning as an example,
enhanced with MeCoD,
object bias entropy increases by 22\% and 27\%,
and P@1 increases by 2\% and 4\% for BERT and RoBERTa, respectively.
This demonstrates that 
alleviating bias contributes to improving the accuracy performance.
Similar results are observed in Preﬁx-tuning.
This demonstrates the generality and effectiveness of our method.
The results of Undersampling-based methods
illustrate that accuracy is sacrificed for debiasing.
The improvement is also reflected in the evaluation results on WIKI-UNI dataset,
as shown in Table \ref{tab: lanka-result-1} 
and Table \ref{tab: lanka-result-2} in Appendix\ref{sec:appendix WIKI-UNI}.


\subsection{Ablation Study}
In order to clarify the source of performance improvement in MeCoD,
we take P-tuning as an example
and conduct ablations by removing particular modules
from MeCoD.
The ablation study results are shown in Table \ref{tab: ablation study}.
We can conclude that 
(1) Object Equalization plays a crucial role in 
alleviating bias, as removing the module causes object bias entropy to decrease.
The decreased accuracy shows that 
Object Equalization module is helpful to improve the accuracy by alleviating the object bias.
(2) Biased Object Obstruction is useful for ensuring accuracy, 
because the contrastive loss forces model not to be affected by biased objects.
But it does not alleviate object bias clearly, which may cause other unexpected problems,
so it's not recommended to be used alone.

\begin{table*}[t]
    \centering
    \tabcolsep=0.19cm
    \begin{tabular}{lccccccc}
      \toprule 
      \multirow{2}{*}{Method} & \multicolumn{7}{c}{Top-k Candidates}\\
      \cmidrule{2-8} 
      & \multicolumn{3}{c}{original prompt }
      &  
      & \multicolumn{3}{c}{subject-masked prompt} \\ 
      \midrule
      \multirow{2}{*}{P-tuning} & \small{Atari(1)} &  \small{\textbf{Boeing(2)}} & \small{IBM(4)}
                                &
                                & \small{IBM(1)} & \small{Atari(2)} & \small{Boeing(22)}  \\
                                & \small{13.789}   &  \small{\textbf{12.980}}    & \small{12.662}
                                & 
                                & \small{11.851} & \small{11.846} & \small{7.071}  \\
      \multirow{2}{*}{P-tuning + MeCoD} &\small{\textbf{Boeing(1)}} & \small{Atari(3)} & \small{IBM(4)}
                                      & 
                                      &\small{Atari(21)} & \small{IBM(27)} & \small{Boeing(53)}\\
                                      &\small{\textbf{13.501}} & \small{11.720} & \small{11.124}
                                      &
                                      &\small{8.510} & \small{8.396} & \small{7.717}\\
      \midrule
      \multirow{2}{*}{Prefix-tuning} & \small{French(1)} &  \small{\textbf{English(2)}} & \small{Russian(3)} 
                                     &
                                     & \small{French(1)} & \small{Russian(2)} & \small{English(5)}  \\
                                     & \small{19.316}   &  \small{18.383}    & \small{18.036}
                                     & 
                                     & \small{16.584} & \small{16.067} & \small{14.894}  \\
      \multirow{2}{*}{Prefix-tuning + MeCoD } & \small{\textbf{English(1)}} &  \small{French(2)} & \small{Russia(3)} 
                                           &   
                                           & \small{Russian(213)} &  \small{French(257)} & \small{English(279)}  \\
                                           & \small{\textbf{20.834}}   &  \small{20.532}   & \small{19.242} 
                                           &
                                           & \small{7.013} & \small{6.526}  & \small{6.269}   \\
    \bottomrule
    \end{tabular}
    \caption{Case Study: 
    Top-k object candidates extracted by original prompt and subject-masked prompt about two fact samples.
    The first (middle row) is the result about ``the developer of B-17 Flying Fortress''.
    The second (bottom row) is the result about ``the native language of Douglas Adams''.
    The bold fonts indicate ground truth,
    the numbers in parentheses are ranks of object candidates
    and the numbers below objects are the corresponding logit values.}
    \label{tab:case study}
\end{table*}

\subsection{Case Study}
\begin{figure}[tbp] 
  \includegraphics[width=\linewidth]{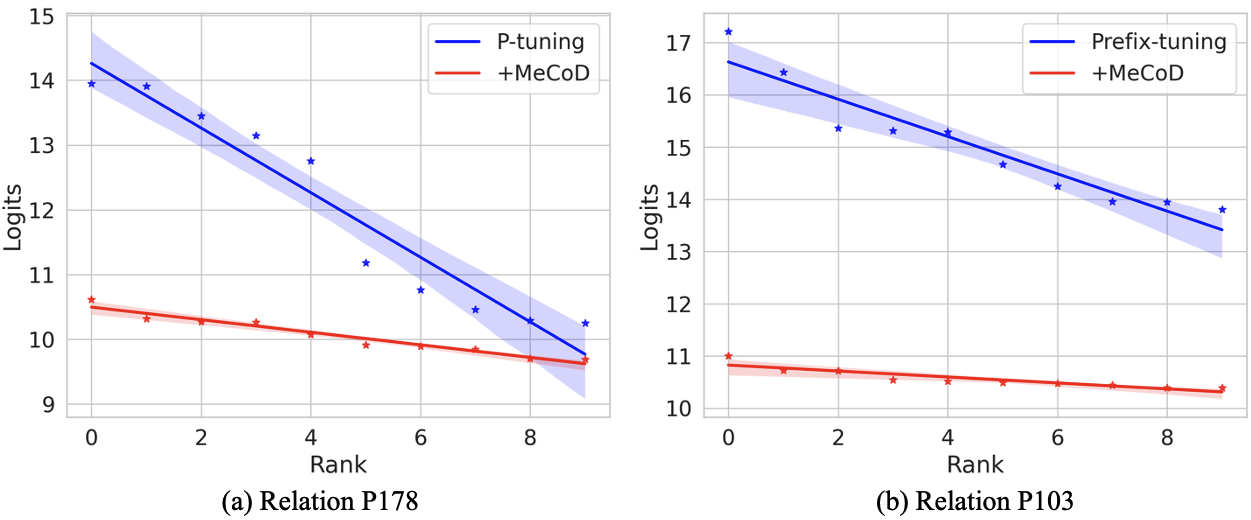}
  \centering
  \caption{Case Study: The regression lines of subject-masked logits.
          (a) shows the result predicted by P-tuning and MeCoD on relation P178.
          (b) shows the result predicted by Prefix-tuning and MeCoD on relation P103.} 
  \label{fig:case study}
\end{figure}

In order to better understand 
how MeCoD contributes to alleviating object bias,
as shown in Figure \ref{fig:case study},
we visualize the regression lines of subject-masked logits 
of P-tuning and Preﬁx-tuning
on relation P178 (developer) and relation P103 (native language), respectively.
The lines corresponding to MeCoD are relatively flatter,
that is, the object bias is alleviated clearly by our method,
which is consistent with quantitative results in Table \ref{tab: result}.
Furthermore,
we investigate the influence of alleviating object bias on 
knowledge extraction by observing
the top-k candidates extracted by original prompt 
and subject-masked prompt respectively on two fact samples, as shown in Table \ref{tab:case study}.
Take P-tuning for example,  
by observing the results extracted by subject-masked prompt,
we find that the candidates' logits of MeCoD are more even,
which means less object bias.
However, P-tuning shows obvious bias on some objects like ``Atari'', ``IBM''.   
Correspondingly,
as for the results extracted by original prompt, 
MeCoD correctly predicts ``Boeing'',
but P-tuning predicts incorrectly on biased object ``Atari'', 
and ranks the correct object at the second place.
Similar results are observed in Prefix-tuning. 
Therefore,
we conclude that object bias is responsible to such incorrect predictions.


\subsection{Discussions}
\begin{table*}[t]
  \centering
  \tabcolsep=0.1cm
  \begin{tabular}{lccccccccc}
    \toprule 
    Prompt token & \multicolumn{9}{c}{MLM candidate words} \\ 
    \midrule
    Front-1      &   . & \textbackslash & ,  & the            & )  & ; & of & and & ? \\
    Front-2      &   . & ,              & of & \textbackslash & -  & ) & ;  & the & ? \\
    Front-3      &   . & of             & ,  & the            & in & ; & The& \textbackslash  & a \\
    \midrule
    Middle-1     &   . & London         & :  & \#\#ville        & Amsterdam
                 &   , & \#\#s            & \textbf{Paris} & - \\
    Middle-2     & Albert & Victor & Max    & \textbf{Paris}   & .  & Robert & Amsterdam  & \#\#s & ? \\
    Middle-3     & Albert & Max    & Victor & Robert  & .  & Eric   & Ann        & Raymond & and\\
    \midrule
    Back-1       &   , & :         & \#\#il & -       & .
                 &   ; & \#\#l     & \#\#el & \#\#lyn \\
    Back-2     & : & \#\#il & -    & ,   & .  & July & August  & operator & ; \\
    Back-3     & . & ;    & ? & !  & |  & ...   & \u0964      & c & -\\
  \bottomrule
  \end{tabular}
  \caption{MLM candidate words of prompt tokens.
           Prompt template used in this case 
           is ``[P][P][P] [MASK] [P][P][P] Claude Arrieu [P][P][P]''.
           [P] indicates prompt token,
           and we use ``Front'', ``Middle'', ``Back'' to
           represent their positions.
           The words in bold indicate ground-truth object.}
  \label{tab:dicussion case-2}
\end{table*}
Our experiments show that object bias of prompt tuning methods 
is undesirable. 
This inspires us to investigate how prompt tuning methods
extract factual knowledge, 
and explore the potential cause of object bias.
Specifically,
we take P-tuning for example and illustrate in the following discussions on
relation P19 (\emph{place-of-birth}).\\
\textbf{Finding Nearest Neighbors.} 
In order to figure out the implication of the prompt token embeddings, 
we follow \citep{Lester2021ThePO} and
find their nearest neighbors 
from the frozen model's vocabulary.
As shown in Table \ref{tab:dicussion case-1}
\footnote{We show Table \ref{tab:dicussion case-1}
in Appendix \ref{sec:appendix Nearest Neighbors}.},
we observe that the prompt tokens in close position 
exhibit similar patterns. 
This indicates
the tokens in different positions 
probably play different roles,
but we can not understand the concrete meaning
by the observation of their nearest neighbors.
Therefore, 
we further analyze the 
the candidate words about prompt tokens,
which is returned by
masked language modeling (MLM) of PLMs. \\
\textbf{Checking MLM Candidate Words.}
Table \ref{tab:dicussion case-2}
illustrates the MLM candidate words of prompt tokens.
Specifically,
the subject is set to ``Claude Arrieu'' 
who was a prolific French composer born in ``Paris''.
Two interesting observations include:
(1) The MLM candidate words of front and back prompt tokens 
are mostly punctuations, while the front also include articles.
This indicates that prompt tuning methods mimic 
human linguistic expression to some extent.
For example, we often start a sentence with the article ``the'' and
end it with the punctuation ``.''.
(2) 
The results of middle prompt tokens exhibit literal correlations
between ``subject'' and ``[MASK]''.
Specifically, it is easy to find that 
the candidate words of Middle-1
are related to the object,
for example ``London'' and  even the ground-truth object ``Paris''.
Note that the Middle-1  is the closest to ``[MASK]''.
As for Middle-3, 
some candidate words can form the names of famous people
with the first token of the subject,
e.g.,  ``Albert Claude'', ``Victor Claude'', ``Robert Claude''.
Note that the Middle-3  is the closest to ``subject''.
The candidate words of Middle-2 can be seen as a mixture of 
the above two type words.
By combining the above two observations,
we conclude that prompt tuning methods
extract factual knowledge
by depending on shallow literal correlations
rather than factual relations.
Furthermore, 
we consider that the shallow correlations 
is one of the potential cause of object bias.
This needs to be demonstrated by rigorous experiments and analysis
in the further works, 
and we just conjecture it intuitively here.
\section{Related work}
\textbf{Language Models as Knowledge Base.}
Since the birth of Pretrained Language Models (PLMs),
researchers have observed that 
there are much knowledge in PLMs.
A typical research line is to explore 
whether the pre-training model can serve as a knowledge base.
\citet{Petroni2019LanguageMA} demonstrated the existence of factual knowledge in PLMs 
and probed the factual knowledge with cloze-style prompts.
Recently, \citet{dai2021knowledge} proposed a knowledge attribution method 
to identify the factual knowledge neurons that store facts in PLMs.
However, 
\citet{cao2021knowledgeable}  and \citet{li2022pre} 
questioned the previous
conclusion by investigating the behaviors of MLMs,
that current MLMs can potentially serve as reliable factual knowledge bases. 
In this paper, 
instead of knowledge storage mechanism of the PLMs,
we mainly investigate the object bias problem 
in factual knowledge extraction
during prompt tuning stage.\\
\textbf{Factual Knowledge Extraction.}
In addition to  manual prompts,
many researchers explored more effective methods for factual knowledge extraction. 
\citet{Jiang2020HowCW} mined prompts through text mining and paraphrasing.
Recently, researchers engage in prompt tuning methods \citep{haviv2021bertese} \citep{Qin2021LearningHT}.
For example,
\citet{Shin2020ElicitingKF} trained a model to generate prompts automatically 
based on gradient optimization.
\citet{Liu2021GPTUT} proposed P-tuning which 
completely abandoned natural language forms
and optimized a continuous prompt for each factual relation. 
\citet{Li2021PrefixTuningOC} proposed a semi-automatic method called Prefix-tuning 
to learn a prefix to add to manual prompts. 
\citet{Newman2022PAdaptersRE} applied Prefix-tuning 
to improve the robustness of factual knowledge extraction.
In this paper,
we observe that prompt tuning methods suffer serious object bias,
and propose a framework MeCoD to alleviate it.

\section{Conclusion}
In this work, 
we position the object bias problem in prompt tuning-based
factual knowledge extraction,
and propose MeCoD, 
a framework for alleviating the
object bias and improving accuracy of factual knowledge extraction.
Experimental results demonstrate
the usefulness and generality of MeCoD.
Besides, 
we argue that 
the shallow association learned by prompt tuning
is one potential cause of object bias.
In the future, 
we are working towards exploring the mechanism behind deriving object bias,
designing more reliable prompt tuning methods for factual knowledge extraction
and investigating the problems on generative pre-trained models,
e.g., GPT2 \citep{Radford2019LanguageMA}, GPT3 \citep{Brown2020LanguageMA}. 

\section*{Limitations}
In this paper,
we focus on masked language models, 
which have been shown very effective and are widely used. 
One limitation of the present study 
is not investigating another representative category of language models,
the generative pre-trained models (e.g., GPT2/3 ( \citet{Radford2019LanguageMA,Brown2020LanguageMA})), 
We leave it for future work.

\section*{Acknowledgments}
We thank the anonymous reviewers for their valuable comments.
This work is supported by the Fundamental Research Funds for Central Universities (No.2022YJS144).

\bibliography{anthology,custom}
\bibliographystyle{acl_natbib}

\clearpage
\appendix
\section{Appendix \label{sec:appendix}}
\subsection{Quantitative Results on WIKI-UNI Datasets \label{sec:appendix WIKI-UNI}}
\begin{table}[h]
  \centering
  \small
  \tabcolsep=0.15cm
  \begin{tabular}{lccc}
    \toprule 
    {Method}  & P@1 & MRR & Entropy\\ 
    \midrule
     LAMA	&14.785  &21.178 	&2.083  \\
    \midrule
     Prefix-tuning &21.806  &28.342  &2.000 \\
     + Undersampling &21.427  &27.846 & 2.150 \\ 
     + MeCoD  &22.745  & 29.479  & \textbf{2.283} \\
     \midrule
     P-tuning  & 22.310 &  29.251&1.776  \\
     + Undersampling  &22.467  &28.982  &1.964  \\
     + MeCoD  &\textbf{22.935}  & \textbf{29.971} & 2.092  \\
    \bottomrule
  \end{tabular}
  \caption{Results on the WIKI-UNI dataset using the BERT-base-cased model, 
  averaged over relations.}
  \label{tab: lanka-result-1}  
\end{table}

\begin{table}[h]
  \centering
  \small
  \tabcolsep=0.15cm
  \begin{tabular}{lccc}
    \toprule 
    Method & P@1 & MRR & Entropy\\ 
    \midrule
     LAMA	& 8.411 & 12.920	& 1.960 \\
     \midrule
     Prefix-tuning &20.536&  27.485&  1.908 \\
     + Undersampling &21.379&  28.163&  2.118  \\ 
     + MeCoD  &\textbf{22.199}&  \textbf{29.0.53}&  \textbf{2.298}   \\
     \midrule
     P-tuning &19.240&  25.934&  1.675  \\
     + Undersampling  &19.351&  25.782&  2.020 \\
     + MeCoD  &20.521  &27.813  & 2.044 \\
    \bottomrule
  \end{tabular}
  \caption{Results on the WIKI-UNI dataset using the RoBERTa-base model,
   averaged over relations.}
  \label{tab: lanka-result-2}  
\end{table}
\subsection{Nearest Neighbors Case \label{sec:appendix Nearest Neighbors}}
\begin{table}[h]
  \centering
  \small
  \tabcolsep=0.08cm
  \begin{tabular}{lcccc}
    \toprule 
    Prompt token & \multicolumn{4}{c}{Nearest words} \\ 
    \midrule
    Front-1      & Discovery & \#\#cam & \#\#final & Kathy \\
    Front-2      & Kathy     & \#\#cam & =         & \#\#cam \\
    Front-3      & =         & :       & \#\#cam   & com \\
    \midrule
    Middle-1     & =         & :       & Kathy     & based\\
    Middle-2     & =         & Kathy   & :         & actress\\
    Middle-3     & =         & actress & suitcase  & divorced\\
    \midrule
    Back-1       & divorced  & Buildings & tells   & suitcase\\
    Back-2       & divorced  & suitcase  & Shannon & psychiatrist\\
    Back-3       & Shannon   & Cheryl    & divorced& interception\\
  \bottomrule
  \end{tabular}
  \caption{Top-4 nearest words of prompt tokens.
           Prompt template used in this case 
           is ``[P][P][P] [MASK] [P][P][P] Claude Arrieu [P][P][P]''.
           [P] indicates prompt token,
           and we use ``Front'', ``Middle'', ``Back'' to
           represent their positions.}
  \label{tab:dicussion case-1}
\end{table}
\end{document}